\documentstyle[sprocl,epsf,epsfig,wrapfig]{article}
\def\PRD{{\em Phys. Rev.} D}
\def\ZPC{{\em Z. Phys.} C}

\def\be{\begin{equation}}
\def\ee{\end{equation}}
\def\bea{\begin{eqnarray}}
\def\eea{\end{eqnarray}}
\def\pyth{{\sc Pythia}}
\def\ari{{\sc Ariadne}}
\def\hwg{{\sc Herwig}}
\newcommand{\etout  }{\mbox{$E_{\rm t,out}$}}

\newcommand{\dkt    }{\mbox{$dk_T^2/(k_T^2+k_0^2)$}}
\newcommand{\dk     }{\mbox{\scriptsize $\!\!\!{dk_T^2/(k_T^2\!+\!k_0^2)}\!\!\!$}}

\begin{document}
\thispagestyle{empty}
\enlargethispage{1in}
\vspace{-2.7cm}
\begin{flushright}
UCL/HEP 97-02\\
NORDITA 97/36P \\
RAL-97-025
\end{flushright}
\vspace{0.7cm}

\title{TUNING MC MODELS TO FIT DIS $\mathbf e\gamma$ SCATTERING EVENTS}

\author{J.A. LAUBER$^{1)}$, L. L\"ONNBLAD$^{2)}$, M.H. SEYMOUR$^{3)}$ }

\address{1) University College London, Gower Street, London WC1E 6BT,
England\\
2) Nordita, Blegdamsvej 17, DK-2100 K\o benhavn, Denmark\\
3) Rutherford Appleton Laboratory, Chilton, Didcot OX11 0QX, England 
}

\maketitle\abstracts{
 Monte Carlo models of DIS e$\gamma$ scattering must describe 
 hard emissions as well as the soft and collinear limits.
 Comparison with the observed experimental
 hadronic energy flow has shown that various 
 models underestimate the high-$p_T$ contribution, in particular at low $x$.
 We have attempted to tune the \hwg , \pyth\ and \ari\
 models to improve the 
 agreement with the data, and to understand the physical implications
 of the changes required.
 Unless the physics of these processes is understood it will
 not be possible to unfold the photon structure functions $F_2^\gamma$
 from the $e^+e^-$ data without large model dependent systematic errors.
}
\section{The Problem}
 The measurement of $F_2^\gamma$ in deep inelastic $e\gamma$ scattering, where
 only one of the electrons is ``tagged'' in the detector and the other
 one escapes unseen,
 involves the determination of the  $\gamma^*\gamma$ invariant mass
 $W$ from the hadronic final state.
 \begin{wrapfigure}[15]{r}{6.4cm}
 \vspace{-0.5cm}
 \epsfig{file=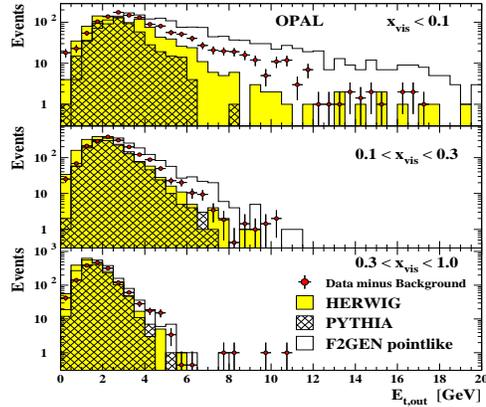,height=6.1cm,width=7cm}
 \vspace{-0.9cm}
 \caption{Transverse energy out of the tag plane.}
 \label{fig:proc_01}
 \end{wrapfigure}
 Because of the non-uniform detection efficiency and incomplete 
 angular coverage the correlation between $W_{\mathrm vis}$ and $W$ critically 
 depends on the modelling of the hadronic final state. 
%
 It has been shown~\cite{OPALPR185} 
 that there exist serious discrepancies in the description 
 of this hadronic final state.
 Fig.~\ref{fig:proc_01} shows the transverse energy out of the plane,
 defined by the tag and the beam. 
 For $x_{\rm vis}>0.1$ all of the generators are adequate, 
 but for $x_{\rm vis}<0.1$ they are
 mutually inconsistent, and in disagreement with the data.  
 At high \etout\ the data show a clear excess over \hwg~\cite{HERWIG}
 and \pyth~\cite{SJO-9401-SJO-9301}, 
 while the pointlike {\sc F2gen}~\cite{BUI-9401} sample exceeds the data.
 Similar discrepancies are 
 observed in the hadronic energy flow per event~\cite{OPALPR185}
 where both \hwg\ and \pyth\ overestimate the energy in the forward
 region ($|\eta|>2.5$) and underestimate the energy in the central region of the 
 detector.\par

\enlargethispage*{-1.0in}
\section{The Tools}

 \begin{wrapfigure}[15]{r}{6.5cm}
 \vspace{-1.5cm}
 \epsfig{file=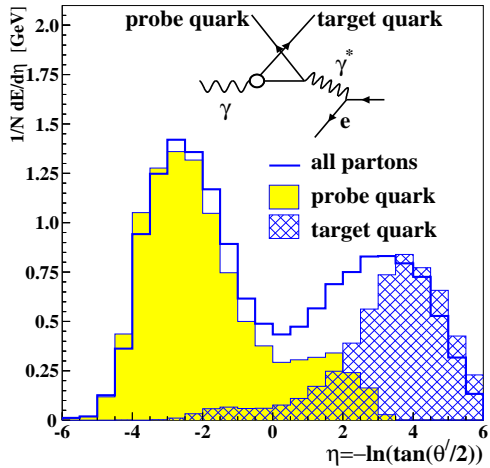,height=6.9cm,width=7cm}
 \vspace{-0.8cm}
 \caption{Energy flows of probe and target quarks.}
 \label{fig:proc_02}
 \end{wrapfigure}

 To study the contributions of the various partons, the \pyth/
 \ari~\cite{Leif}
 energy flow in the lab frame as a function of pseudorapidity 
 is plotted  in Figure~\ref{fig:proc_02} 
 for the quark that couples to the off-shell
 probe photon $\gamma^\star$, denoted the probe quark, and for the
 quark that couples to the quasi-real target photon $\gamma$, denoted
 the target quark. The total energy flow
 of all partons after gluon radiation is also shown. The direction
 of the tagged electron is always at negative $\eta$.
 It is apparent that the hump at negative $\eta$ stems mostly from the
 probe quark which is scattered in the hemisphere of the tag, while the
 hump at positive $\eta$ originates mostly from the target quark
 in the opposite hemisphere of the struck photon.

\begin{figure}[t]
 \vspace{-0.6cm}
 \hspace{-0.cm}
 \begin{minipage}[t]{0.32\linewidth}
   {\epsfig{file=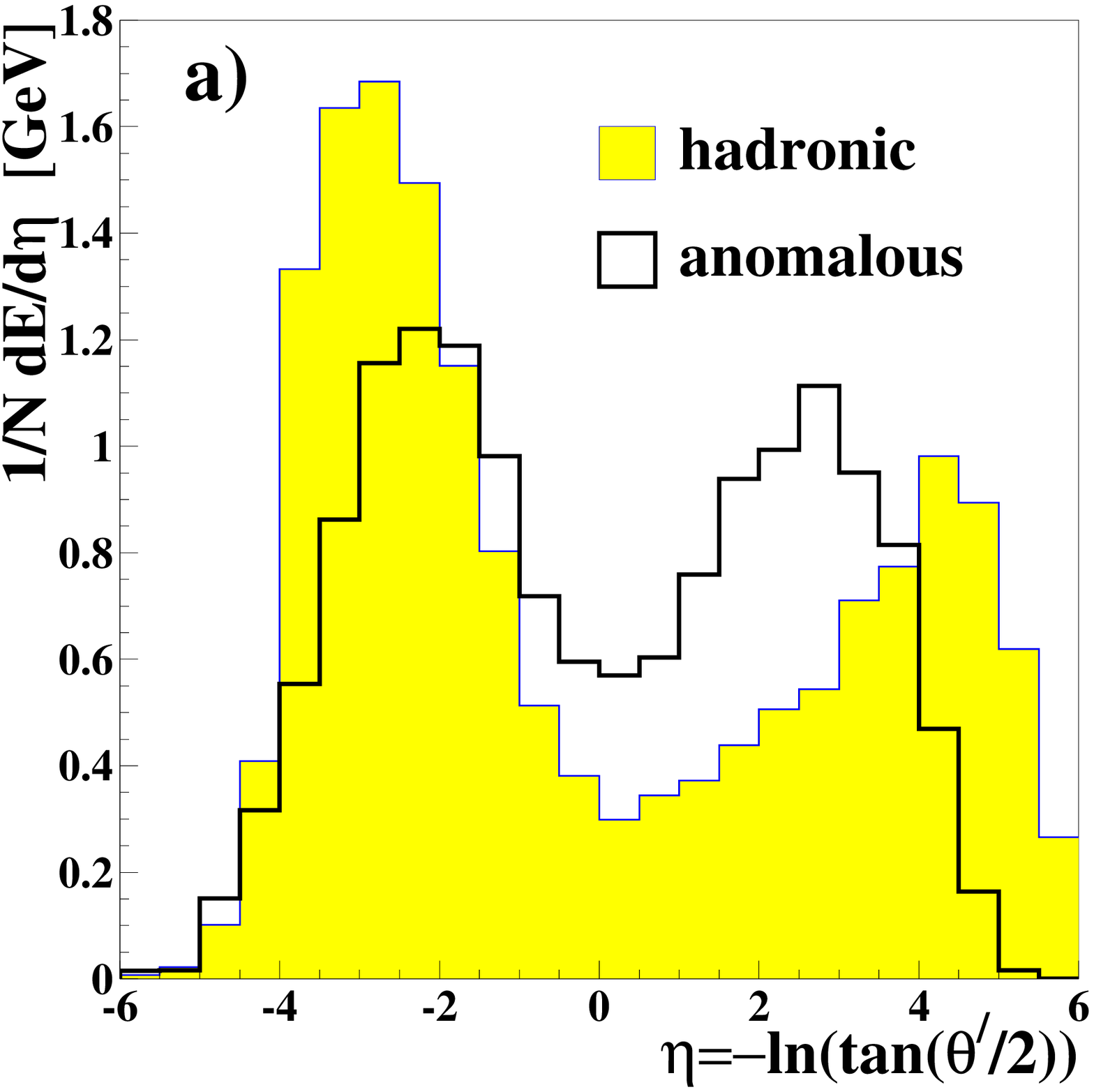,height=4.1cm,width=1.05\linewidth}}
 \end{minipage}
 \begin{minipage}[t]{0.32\linewidth}
   {\epsfig{file=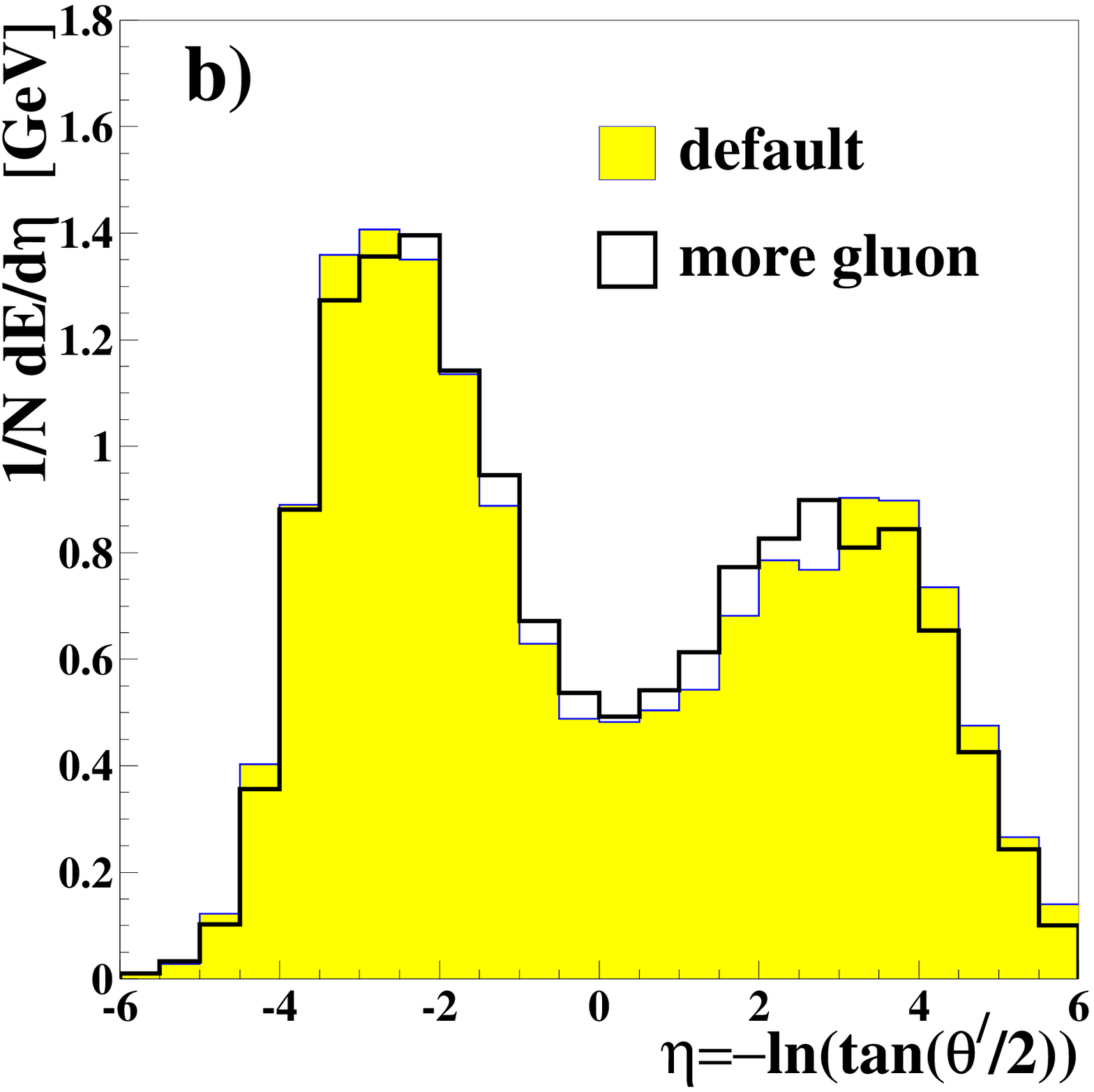,height=4.1cm,width=1.05\linewidth}}
 \end{minipage}
 \begin{minipage}[t]{0.32\linewidth}
   {\epsfig{file=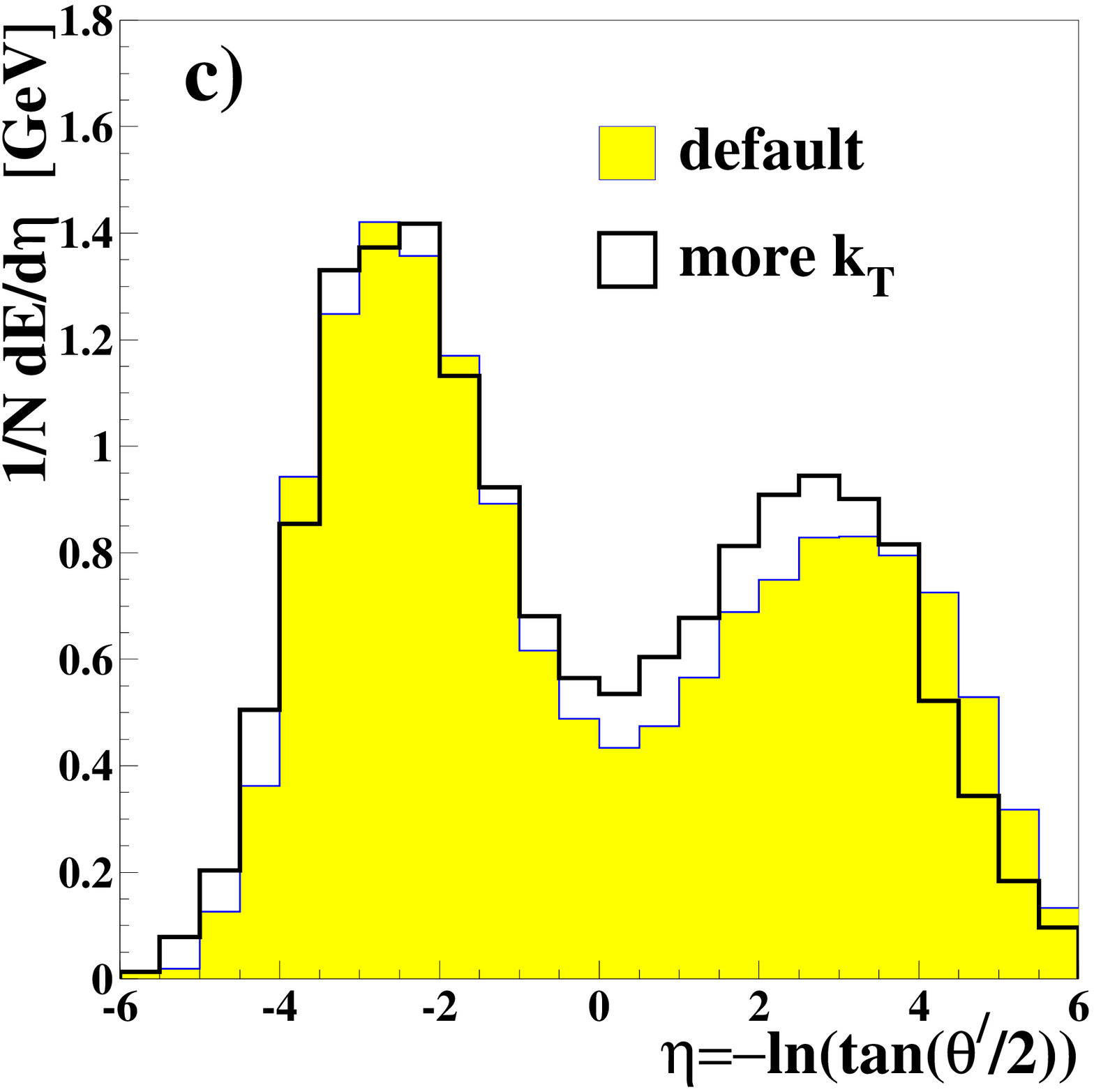,height=4.1cm,width=1.05\linewidth}}
 \end{minipage}
  \caption{\label{fig:proc_03} Different methods of increasing the
  energy flow in the central detector region using the \pyth/\ari\
  Monte Carlo generators.}
 \hspace{-0.2cm}
\end{figure}

 From comparisons of the hadronic energy flow of the data with the
 various models, it became apparent that the energy flow of the probe
 quark needs to be shifted to lower $\eta$, corresponding to an 
 increased transverse energy. This can be achieved in
 several ways:

 Anomalous events carry more transverse momentum than hadronic events.
 Figure~\ref{fig:proc_03}a) shows the partonic energy flow 
 after gluon radiation.  Increasing the fraction of 
 anomalous to hadronic or VMD type events would have the desired effect,
 but the parton density functions (PDF) used (in this case 
 SaS1D~\cite{SCH-9501} in \pyth\ and GRV~\cite{GLU-9201-GLU-9202} 
 in \hwg) does not readily allow changing this ratio.

 Another way to increase the transverse energy is to allow for more
 gluon radiation. This can be achieved by augmenting the inverse
 transverse size of the remnant, $\mu$, in the \ari\
 colour dipole model,
 shown in  Figure~\ref{fig:proc_03}b). $\mu$ is set proportional to
 the intrinsic $k_T$ of the
 struck quark on an event-by-event basis.  For VMD events, 
 $k_T$ is gaussian with a width of 0.5 GeV.  For anomalous events $k_T$
 follows a power law.  But even a generous increase of the $\mu$ 
 parameter ($\mu=10$) 
 has a relatively small effect on the partonic energy flow.

 Increasing the intrinsic transverse momentum $k_T$ of the struck photon
 is another way of directly influencing the angular distribution of the
 hadronic final state.   Figure~\ref{fig:proc_03}c) shows the energy flow
 for \pyth\ events with default settings and enhanced $k_T$.
 This appears to be the most promising method.

\section{The Fix}

 \begin{wrapfigure}[10]{r}{4.9cm}
 \vspace{-1.9cm}
 \epsfig{file=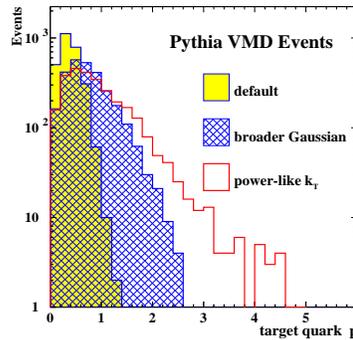,height=5.cm,width=5.1cm}
 \vspace{-0.9cm}
 \caption{\label{fig:proc_04} Transverse momentum of the target quark
                              in \pyth/\ari .}
 \end{wrapfigure}

 In \pyth\ the PDF determines whether an event is generated as a VMD
 or an anomalous event.
 The intrinsic $k_T$ of the quasi-real 
 photon can be controlled with parameters~\cite{SJO-9401-SJO-9301}.
 Just increasing the width of the gaussian distribution does not produce
 events that populate the region of high \etout\ at low $x$ observed
 in the data (Fig.~\ref{fig:proc_01}).  A similar deficiency had been
 observed in the resolved photoproduction data at Zeus~\cite{Zeus-B354}, 
 which lead to the introduction of a
 power-like $k_T$-distribution of the form \dkt ,
 improving the distributions of the photon remnant. $k_0$ is a constant,
 for which 0.66 GeV was used.
 The \pyth\ parameters only allow adjusting the $k_T$ for VMD type events,
 figure~\ref{fig:proc_04}, but not for anomalous evens.
 To change the intrinsic $k_T$ of anomalous events a
 gaussian smearing is added in quadrature~\cite{Torbjorn}.

\begin{figure}
\vspace{-0.5cm}
 \begin{minipage}[r]{0.45\linewidth}
  \epsfig{file=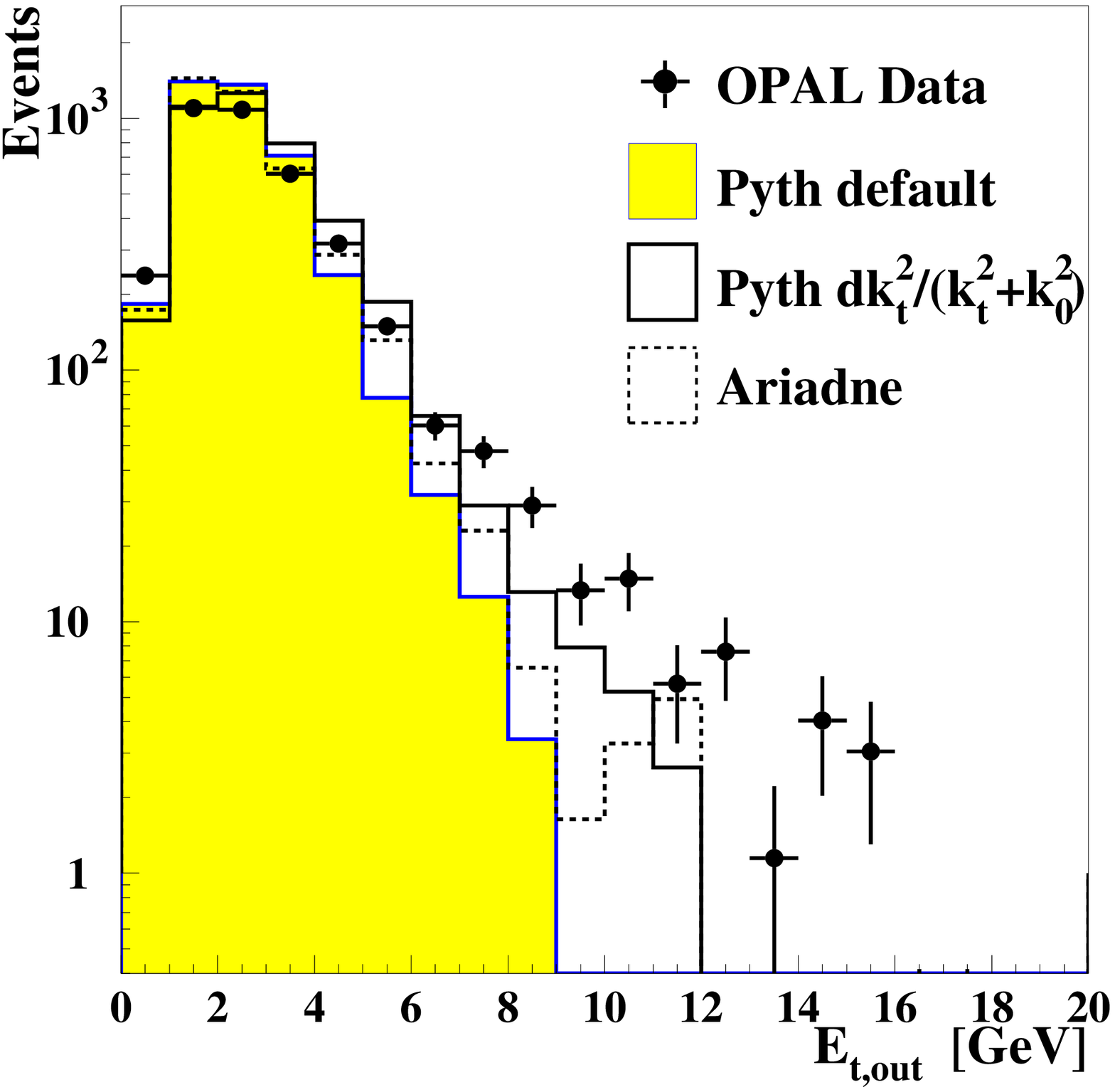,height=5.1cm,width=5.5cm}
  \vspace{-0.5cm}
  \caption{\label{fig:proc_05} \pyth : 
     Transverse energy out of the tag plane.}
 \end{minipage}\hfill  
 \begin{minipage}[r]{0.45\linewidth}
  \epsfig{file=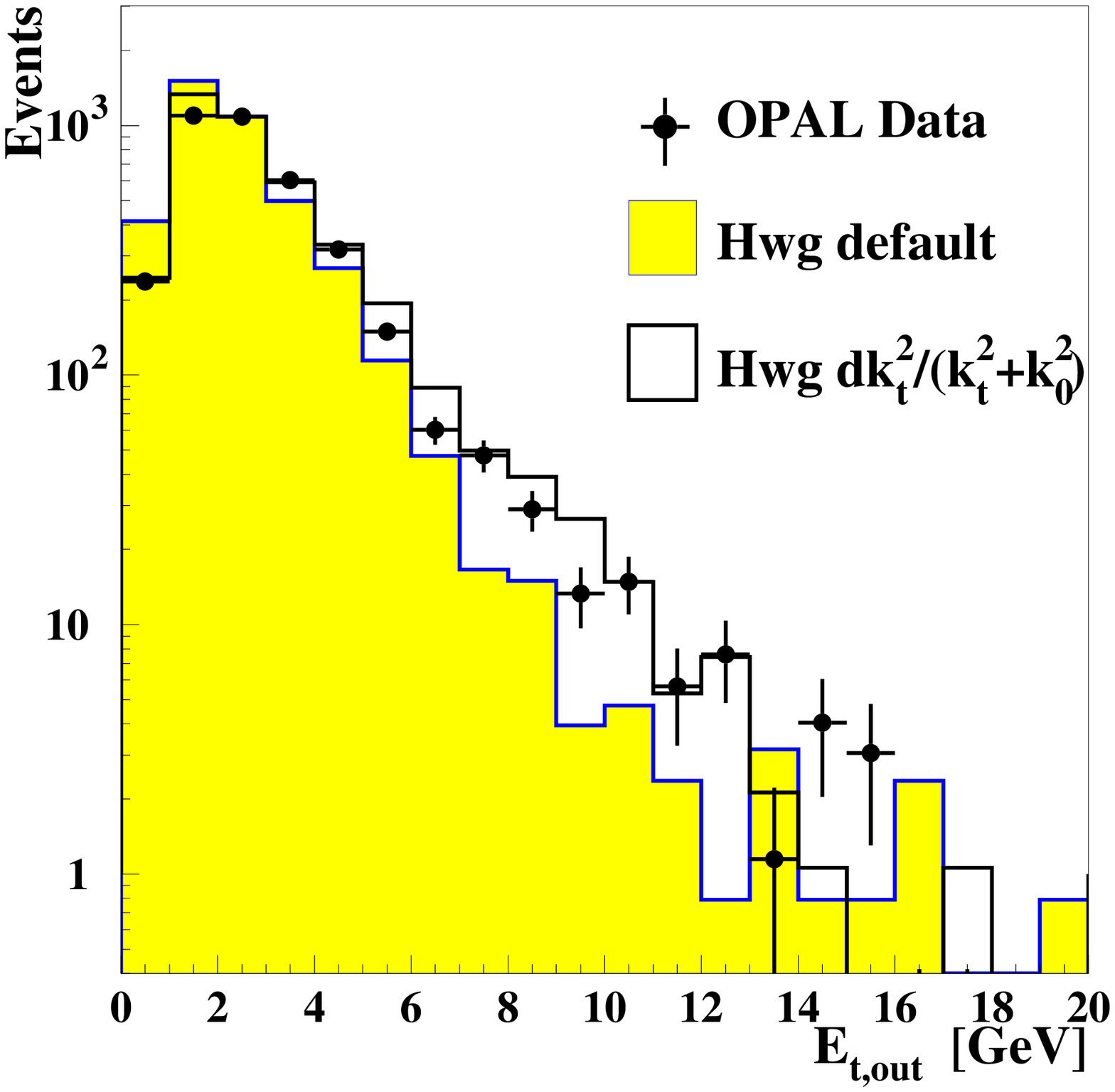,height=5.1cm,width=5.5cm}
  \vspace{-0.5cm}
  \caption{\label{fig:proc_06} \hwg : 
     Transverse energy out of the tag plane.}
 \end{minipage}  
\end{figure}

\begin{figure}
 \vspace{-0.2cm}
 \begin{minipage}[r]{0.45\linewidth}
  \epsfig{file=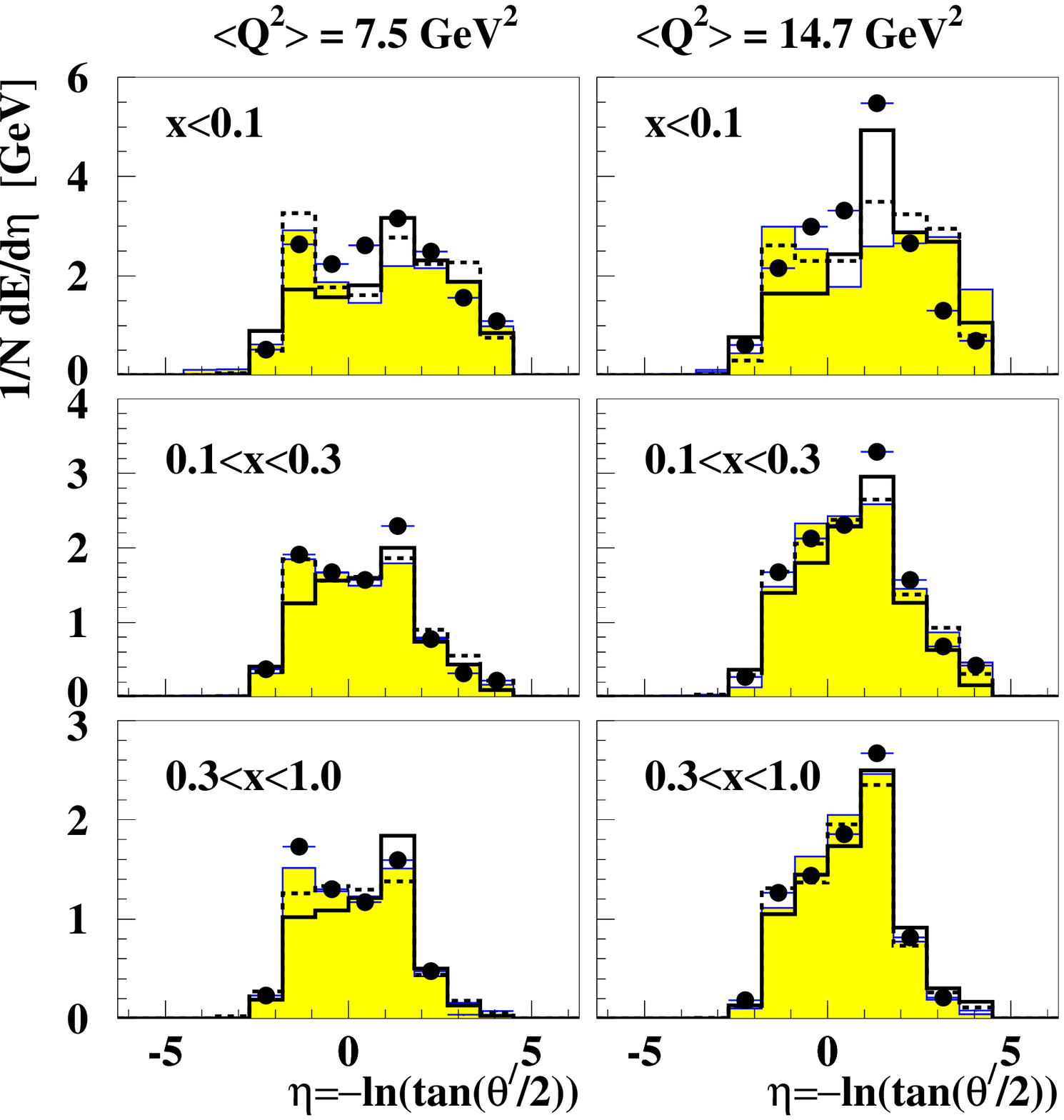,height=6.5cm}
  \vspace{-0.5cm}
  \caption{\label{fig:proc_07} \pyth : 
     Hadronic energy flow as function of $x$ and $Q^2$.
     The symbols are the same as in Fig. 5}
 \end{minipage}  \hspace{0.5cm}
 \begin{minipage}[r]{0.45\linewidth}
  \epsfig{file=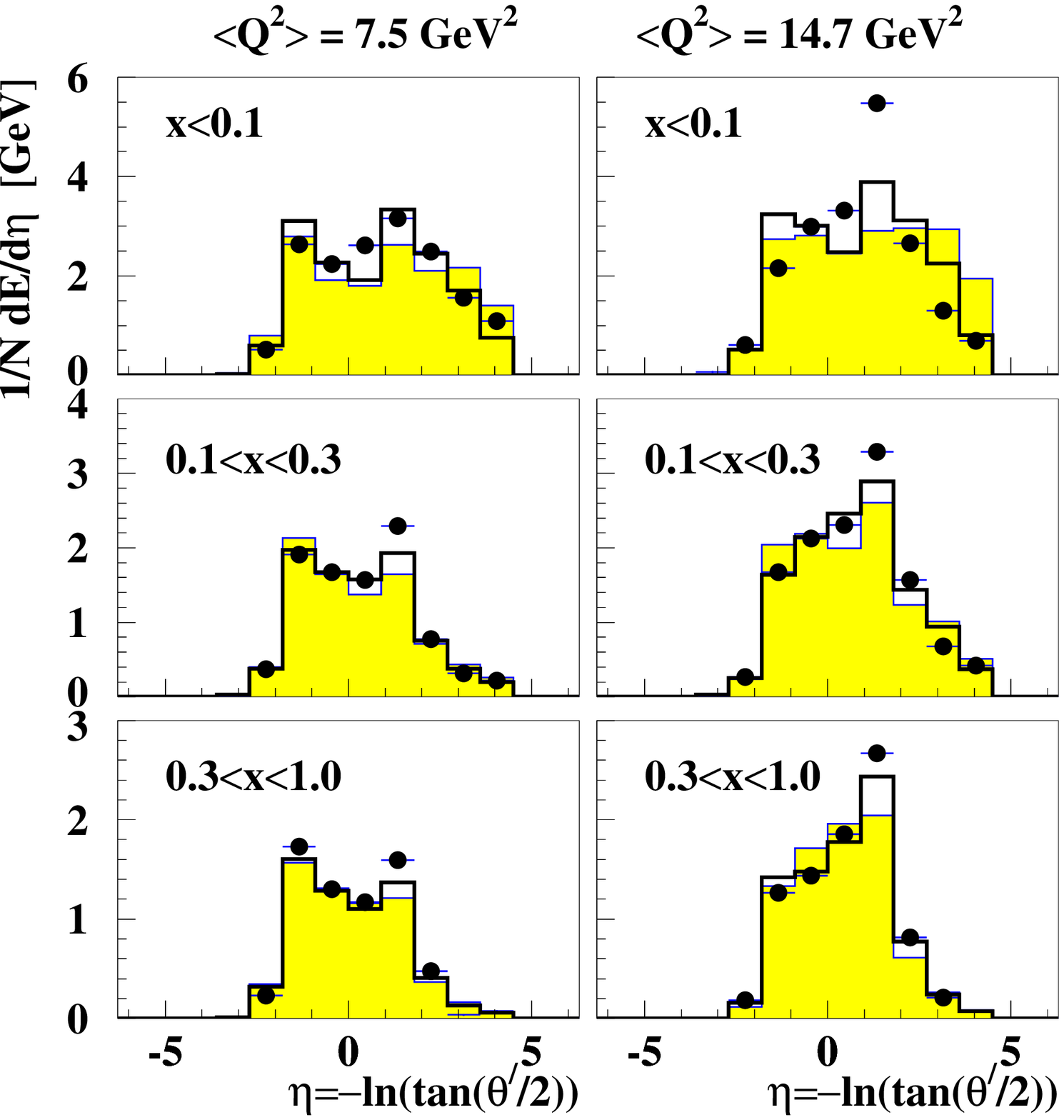,height=6.5cm}
  \vspace{-0.5cm}
  \caption{\label{fig:proc_08} \hwg :  
     Hadronic energy flow as function of $x$ and $Q^2$.
     The symbols are the same as in Fig. 6}
 \end{minipage}  
\end{figure}

 Figure~\ref{fig:proc_05} show the \etout\ and figure~\ref{fig:proc_07} 
 the hadronic energy flows on detector level of \pyth\ with default
 parameter settings and with the \dkt\ distribution for VMD
 plus a gaussian smearing
 of the anomalous events, compared to the OPAL data taken 
 in 1993--1995 at $\sqrt{s_{ee}}=91$ GeV. In addition the \ari\ distributions
 with enhanced gluon radiation are shown.   While the \etout\ spectrum
 has been improved, it still falls short of the data in the tail of 
 the distribution.  The hadronic energy flow generated by the enhanced
 \pyth\ recreates the peak on the remnant side (positive $\eta$) seen
 in the data at low $x_{\rm vis}$, at the expense of a somewhat worse fit
 on the tag side.

 The 2-jet rates, listed in Table~\ref{tab:jetpyth},
 found with the cone algorithm, requiring a minimum
 $E_{T,{\mathrm jet}}>3$ GeV in the pseudorapidity range of 
 $|\eta_{\mathrm jet}|<2$ are almost 
 doubled over the default version of \pyth , but are still substantially
 lower than the data~\cite{TTT}.

\nopagebreak
\begin{table}
 \hspace{-0.2cm}
 \begin{minipage}[t]{0.45\linewidth}
     \begin{tabular}{|l||r|r|r|}
        \hline
                &  0 jet &  1 jet &  2jet \\ \hline
        Data    & 30.7\% & 63.8\% &  5.4\% \\ \hline
        \pyth   & 32.8\% & 65.5\% &  1.7\% \\ \hline
        \dk     & 36.5\% & 60.6\% &  2.9\% \\ \hline

    \end{tabular}
    \caption{Jet rates for \pyth}
    \label{tab:jetpyth}
  \end{minipage}\hspace{0.8cm}
  \begin{minipage}[t]{0.45\linewidth}
    \begin{tabular}{|l||r|r|r|}
        \hline
                &  0 jet &  1 jet &  2jet \\ \hline
        Data    & 30.7\% & 63.8\% &  5.4\% \\ \hline
        \hwg    & 34.0\% & 63.6\% &  2.4\% \\ \hline
        \dk     & 31.0\% & 63.0\% &  5.9\% \\ \hline
    \end{tabular}
    \caption{Jet rates for \hwg}
    \label{tab:jethwg}
  \end{minipage}
  \vspace{-0.2cm}
\end{table}


 \hwg\ separates events dynamically into hadronic and anomalous type.
 A similar \dkt\ distribution of the intrinsic transverse momentum
 of the photon can be added by hand.  The results of this are
 shown in figures~\ref{fig:proc_06} and \ref{fig:proc_08}.  Both the
 \etout\ and the energy flows are greatly improved
 with the inclusion of the power-like $k_T$ distribution, with the
 exception of the peak in the energy flow at low $x_{\rm vis}$ -- high
 $Q^2$, which still falls short of the data.

 The cone algorithm 2-jet rate, listed in Table~\ref{tab:jethwg},
 is more than doubled over the default version of \hwg\ and is in 
 agreement with the data.

\section{Conclusion}

 The power-like distribution of the intrinsic transverse momentum
 of the struck photon of the form \dkt\ greatly improves the hadronic
 final state distributions of both \pyth\ and \hwg . This improved 
 description of the data should reduce the model-dependent systematic
 errors in the unfolded result of the photon structure function 
 $F_{2}^{\gamma}$. More fine-tuning is required.

 It should be stressed, though, that this is just an {\it ad hoc}
 solution which does not explain the origin of the discrepancies.
 It appears that the photon displays a more pointlike behaviour at low
 $x$ than predicted. $F_{2}^{\gamma}$ and the $\gamma^\star\gamma$ 
 fragmentation are not orthogonal.  
 Changing the latter will affect the measurement of the
 former.  A thorough enquiry is urgently needed.


\vfill\eject

\end{document}